\documentclass[sigconf,authorversion]{acmart}
\makeatletter
\renewcommand\@formatdoi[1]{\ignorespaces}
\makeatother
\usepackage[noend]{algorithmic}
\usepackage{algorithm}
\usepackage{multirow}
\usepackage{flushend}
\usepackage{amsmath,amssymb}
\usepackage{mathtools}
\usepackage[acronym]{glossaries}
\usepackage{units}
\usepackage{booktabs}
\usepackage[inline]{enumitem} 

\setcopyright{none}

\acmConference[REVEAL '19]{the ACM RecSys Workshop on Reinforcement Learning and Robust Estimators for Recommendation Systems}{September 20th, 2019}{Copenhagen, Denmark}
\acmYear{2019}
\copyrightyear{2019}

\begin{document}
\title[Bandit Feedback for Offline Recommender System Evaluation]{On the Value of Bandit Feedback for\\ Offline Recommender System Evaluation}

\author{Olivier Jeunen}
\affiliation{
  \institution{University of Antwerp}
  \city{Antwerp}
  \country{Belgium}
}
\email{olivier.jeunen@uantwerp.be}

\author{David Rohde}
\affiliation{
  \institution{Criteo AI Lab}
  \city{Paris}
  \country{France}
}
\email{d.rohde@criteo.com}

\author{Flavian Vasile}
\affiliation{
  \institution{Criteo AI Lab}
  \city{Paris}
  \country{France}
}
\email{f.vasile@criteo.com}

\renewcommand{\shortauthors}{O. Jeunen et al.}

\begin{abstract}
In academic literature, recommender systems are often evaluated on the task of next-item prediction.
The procedure aims to give an answer to the question: ``Given the natural sequence of user-item interactions up to time $t$, can we predict which item the user will interact with at time $t+1$?''.
Evaluation results obtained through said methodology are then used as a proxy to predict which system will perform better in an online setting.
The online setting, however, poses a subtly different question: ``Given the natural sequence of user-item interactions up to time $t$, can we get the user to interact with a recommended item at time $t+1$?''.
From a causal perspective, the system performs an intervention, and we want to measure its effect.
Next-item prediction is often used as a fall-back objective when information about interventions and their effects (shown recommendations and whether they received a click) is unavailable.

When this type of data \emph{is} available, however, it can provide great value for reliably estimating online recommender system performance.
Through a series of simulated experiments with the RecoGym environment, we show where traditional offline evaluation schemes fall short.
Additionally, we show how so-called \emph{bandit feedback} can be exploited for effective offline evaluation that more accurately reflects online performance.
\end{abstract}

%
%

\maketitle

\section{Introduction}
Traditionally, recommender systems are evaluated either through offline methods, online methods, or user studies~\cite{Herlocker2004,Shani2011,Aggarwal2016}.
In the offline setting, a previously collected dataset of preference expressions (be it explicit, implicit, or logged feedback) is used to assess the performance of competing recommendation methods.
Online experiments deploy every competing method to a portion of real user traffic, and users' interactions with the systems are subsequently measured (often through A/B-tests, inter- or multi-leaving~\cite{Kohavi2009,Chapelle2012,Brost2016}).
As online methods require a large number of resources and time, they are more costly than their offline counterpart~\cite{Gilotte2018}.
Finally, the most expensive option, user studies are small-scale analyses where users' interactions with the system are studied in a more detailed manner, usually followed by qualitative questionnaires.
In mixed-methods research, multiple of these variants are combined~\cite{Garcia-Gathright2018}.
Because of the costly nature of these latter options, offline evaluation methodologies often remain a necessity when assessing algorithmic performance.
However, results obtained through traditional offline evaluation schemes are often poorly correlated with true online performance~\cite{Beel2013,Garcin2014,Rossetti2016}.
A large portion of the academic literature surrounding recommender systems utilises offline evaluation procedures stemming from the broader field of supervised learning.
In this setting, all true labels are assumed to be known and techniques like bootstrapping or $k$-fold cross-validation have been shown to provide accurate performance estimates~\cite{Efron1982,Stone1974,Kohavi1995}.
In a recommender systems context, this line of research focuses on \emph{organic} user behaviour: trying to find the items that naturally complement an already existing user sequence.

The recommender systems use case, however, is in some ways more closely related to reinforcement learning~\cite{Sutton1998}, multi-armed~\cite{Robbins1952} and contextual bandits~\cite{Langford2008}.
Here, the true \emph{labels} or \emph{rewards} are for the most part unknown.
We observe rewards only for the actions (\emph{recommendations}) that were actually performed (\emph{shown to the user}).
This is known as the \emph{bandit feedback} setting~\cite{Bottou2013,Swaminathan2015,Joachims2018}.
Stemming from the reinforcement learning field of off-policy or counterfactual evaluation, a large body of work has recently focused on applying these techniques to provide accurate offline estimators of online recommender performance~\cite{Agarwal2017,Gilotte2018,Gruson2019,Chen2019,Jagerman2019}.
This second line of research aims to perform interventions (recommendations) that influence the user in some optimal way (leading them to click on, or purchase an item).

In this work, we present a comparison study of techniques from both fields.
We investigate the value of using bandit feedback for recommender system evaluation, and show where traditional techniques (focusing solely on organic feedback) fall short.
Through a range of experiments with the RecoGym environment~\cite{Rohde2018}, we empirically validate our findings.

\section{Methodology}
$k$-fold leave-one-out cross-validation (LOOCV) is one of the most used offline evaluation schemes in the literature.
For every user sequence, an item is (or multiple items are) randomly sampled to be part of the test set.
What remains of the user sequences then makes up the training set.
Based on the training set, every model then generates a set of top-$N$ recommendations for every user.
Algorithms that can rank the missing sampled items highest in the set of recommendations are then assumed to be the best performers in an online environment as well.
In order to provide a robust estimate, this process is averaged over $k$ different runs.
As this technique has been used widely and recently to present new models as the state of the art~\cite{Ning2011,Christakopoulou2016,Li2017,He2017,Ning2017,Otunba2017,JYang2018,Zhang2018,Christakopoulou2018}, we adopt it as the representative for traditional offline evaluation in our experiments.
Temporal evaluation procedures that take the chronological ordering of user-item interactions into account have recently gained traction as well~\cite{Jugovac2018,JeunenDS2019}, but are out-of-scope for the purposes of this paper.
The RecoGym environment simulates users' interests changing over time, but items remain stationary.
As such, experiments showed no significant difference between a random or time-based split.
We will consider hit-rate-at-1 (HR@1) as evaluation metric: the ratio of correct item predictions over all users.
In this context, HR@1 is identical to Precision- and Recall@1.

Counterfactual or off-policy evaluation methods often use estimators based on importance sampling or inverse propensity scoring (IPS)~\cite{Owen2013}.
By re-weighting (user, action, reward)-triplets according to how likely they are to occur under a new policy as compared to the old policy, various estimators for the performance of the new policy can be derived.
The Clipped IPS (CIPS) estimator is an extension to classical IPS, exchanging variance for a pessimistic bias by putting a hard upper bound on these weights~\cite{Ionides2008,Bottou2013,Gilotte2018}.
Assume we have a dataset $\mathcal{D}$ consisting of $n$ logs $(x_i,a_i,p_i,\delta_i)$, where $x_i \in \mathbb{R}^d$ describes the user state, $a_i$ is an identifier representing the action that was taken, $p_i \in (0,1)$ denotes the probability with which that action was taken by the logging policy, and $\delta_i \in \{0,1\}$ is the observed reward.
Now, the CIPS estimator for a new policy $\pi$ can be computed on samples from $\mathcal{D}$ as shown in Equation~\ref{eq:1}, where $M$ denotes the maximally allowed sample weight.
\begin{equation}\label{eq:1}
    \text{CIPS}(\pi, \mathcal{D}) = \frac{1}{n}\sum_{i = 1}^{n}\delta_i\cdot\min\left(M,\frac{\pi(a_i|x_i)}{p_i}\right)
\end{equation}
When the rewards $\delta_i$ are clicks, Equation~\ref{eq:1} provides an estimate of the click-through-rate (CTR) that a new policy $\pi$ will generate when deployed.
Note that the logging policy $\pi_0$ needs to be stochastic, and have support over the same actions as $\pi$.
The target policy $\pi$, however, can be deterministic.

\section{Experimental Results}
We compare the traditional and counterfactual evaluation procedures as laid out in the previous sections with results obtained through a simulated A/B test\footnote{A notebook with all source code can be found at: \url{https://git.io/fjyYq}}.
Six recommendation approaches are compared: a random baseline, a popularity baseline recommending the item with the most organic views, a personalised popularity baseline recommending the item the specific user has organically viewed most often in the past, a latent-factor model based on a singular-value-decomposition of the user-item matrix~\cite{Zhang2005}, an item-based k-nearest-neighbours model~\cite{Sarwar2001}, and a user-based k-nearest-neighbours model~\cite{Herlocker1999}.
For illustrative purposes and brevity, we include only traditional methods in our comparison.
Our findings, however, are model-agnostic and general.
All models were trained on logged organic feedback obtained through the RecoGym environment with $2\,000$ users and $2\,000$ items.
This leads to roughly $40\,000$ organic user-item interactions and $160\,000$ bandit-feedback samples.
Note that all considered models only use organic information to generate recommendations: no bandit feedback is taken into account at learning time.
We test our models on a set of $5\,000$ unseen users and the same set of items; with $100\,000$ organic and $390\,000$ bandit samples.
The logging policy when generating test samples was stochastic and based on the personalised popularity baseline.

\begin{figure}
    \centering
    \includegraphics[scale = .57, trim = {3mm 5mm 0 5mm}]{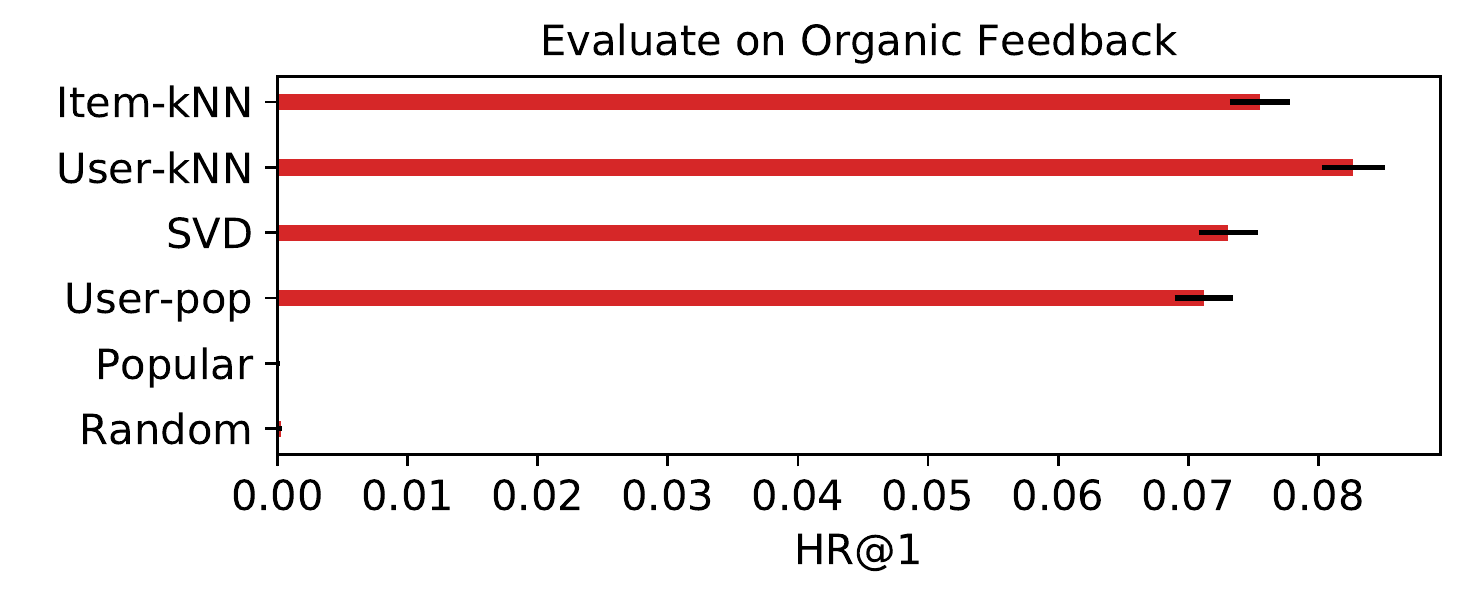}
    \caption{HR@1 as measured through 10-fold LOOCV on a logged dataset with organic feedback.}
    \label{fig:1}
\end{figure}

\begin{figure}
    \centering
    \includegraphics[scale = .57, trim = {3mm 5mm 0 8mm}]{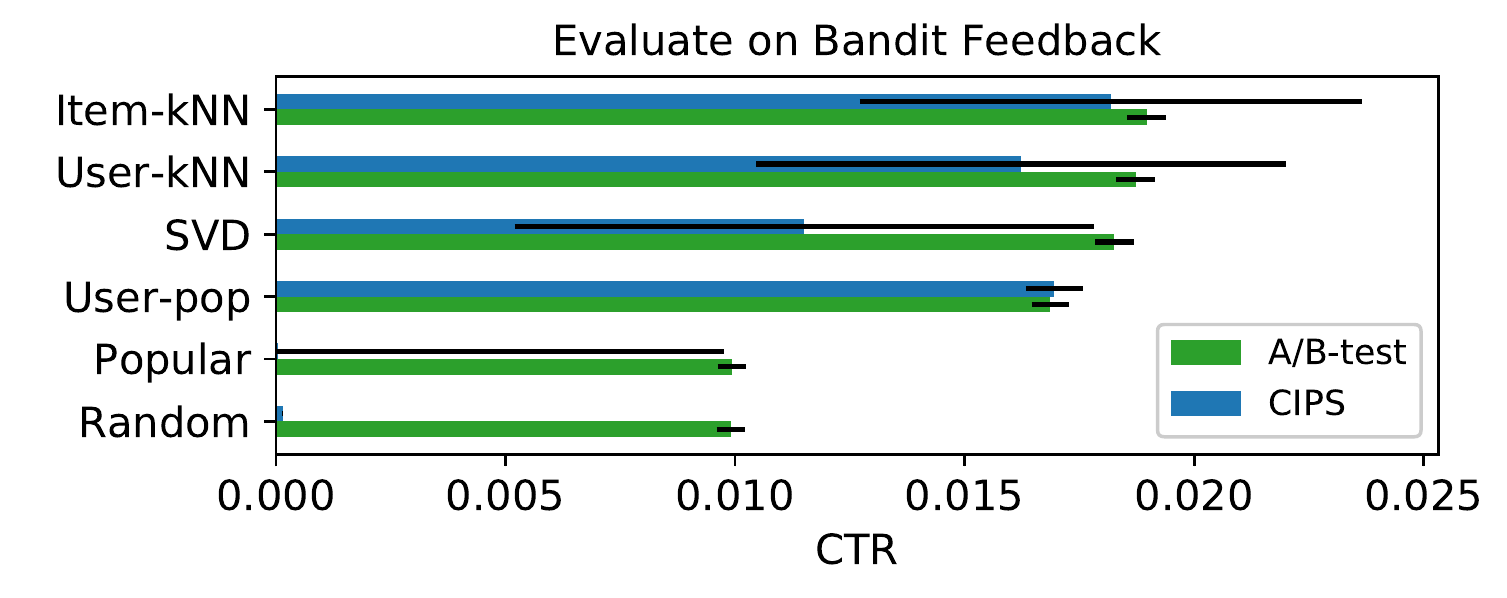}
    \caption{CTR estimate through the CIPS estimator on a logged dataset with bandit feedback ($M=15$), and as measured through a simulated A/B-test.}
    \label{fig:2}
\end{figure}

Figure~\ref{fig:1} shows the achieved HR@1 for all models, obtained through 10-fold LOOCV on the logged organic feedback.
Figure~\ref{fig:2} shows the estimated CTR for all models, obtained through the CIPS estimator on the logged bandit feedback, as well as measured results from a simulated A/B test.
Key observations from these results are as follows:
\begin{enumerate*}
    \item LOOCV generates wildly different results in terms of absolute values, ratios and rankings among competing algorithms. These findings are in line with those presented in~\cite{Jeunen2018}.
    \item The counterfactual CIPS estimator succeeds in providing sensible confidence intervals for the CTR. Although these intervals are wide, their true CTR value is almost always captured\footnote{See~\cite{Maurer2009,Bottou2013} for additional discussion regarding the -sometimes poor- coverage of traditional confidence intervals for importance sampling estimators.}. When ranking competing algorithms according to their upper confidence bound, we are able to infer the true ranking as obtained through the A/B test.
    \item Due to an insufficient sample size, CIPS fails to accurately predict online performance for the Random baseline.
\end{enumerate*}

\section{Conclusion}\label{sec:conclusion}
We presented an overview of the most often-used evaluation procedures for recommender systems, along with the distinction between \emph{organic} and \emph{bandit} feedback and how these terms relate to \emph{supervised} and \emph{counterfactual} evaluation techniques.
Through a series of simulated experiments with RecoGym, we showed that algorithmic performance on predicting organic user behaviour is not necessarily a good proxy for the bandit task.
When properly tuned, counterfactual estimators such as clipped IPS can accurately represent model utility in an online setting.
As such, when bandit feedback is available, it can be exploited for more effective evaluation.

\bibliographystyle{ACM-Reference-Format}
\bibliography{bibliography}

\end{document}